\documentstyle[12pt,epsfig,epsf]{article}
\def\lsim{\raise0.3ex\hbox{$<$\kern-0.75em\raise-1.1ex\hbox{$\sim$}}}
\def\gsim{\raise0.3ex\hbox{$>$\kern-0.75em\raise-1.1ex\hbox{$\sim$}}}
\def\be{\begin{equation}}
\def\ee{\end{equation}}
\def\ba{\begin{eqnarray}}
\def\ea{\end{eqnarray}}
\def\bea{\begin{eqnarray}}
\def\eea{\end{eqnarray}}
 2
\def\m0{m_{D0}}

\newcommand{\tr}{{\rm Tr}}
\def\d{\partial}

\begin{document}

\title{Magnetic Screening in Hot Non-Abelian Gauge Theory}
\author{
A. Cucchieri, F. Karsch and P. Petreczky\\
Fakult\"at f\"ur Physik, Universit\"at Bielefeld,\\
\qquad  P.O. Box 100131, D-33501 Bielefeld, Germany\\
}
\date{\today}
\maketitle
\begin{abstract}\noindent
We analyze the large distance and low-momentum behavior of the magnetic 
gluon propagator of the SU(2) gauge theory at finite temperature.
Lattice calculations within the $4$-dimensional as well as the effective, 
dimensionally reduced $3$-dimensional gauge theories in generalized Landau gauges
and MAG show that the magnetic propagator 
is strongly infrared suppressed in Landau gauges but stays large and finite
in MAG. Despite these differences in the low-momentum
behavior of the propagator calculated in different gauges the magnetic
fields are exponentially screened in all gauges considered. 
From the propagator calculated in maximally Abelian gauge we find
for the screening mass, $m_M = (1.48\pm 0.17) T$ at $T=2\;T_c$. 
\end{abstract}

{PACS numbers: 11.10.Wx, 11.15.Ha, 12.38.Mh}
\vskip0.4truecm

Magnetic fields in hot non-abelian gauge theories have to be screened at 
high temperature for any perturbative description of the plasma phase   
to make sense \cite{Linde}. In fact,
there is plenty of evidence from analytic \cite{gap,nair} as well as numerical 
\cite{heller95,karsch98}
calculations for the screening of electric and magnetic fields in non-abelian 
gauge theories at high temperature. The way this is realized on the partonic
level, however, is poorly understood. Our intuitive understanding of screening
at high temperature is largely influenced by the perturbative concept
of electric and magnetic screening masses \cite{gross81}. However, while the
former is calculable at least to leading order in perturbation theory the latter
is intrinsically non-perturbative.  Its existence has been postulated to 
render the perturbative expansion finite \cite{Linde}. 

In the vacuum the gluon propagator in Landau gauge is expected to be
infrared suppressed. 
The existing rigorous bounds on the unrenormalized gluon propagator in 
momentum space imply that
it is less singular than $p^{-2}$ in $4$ dimension and $p^{-1}$ in $3$ dimension
\cite{zwanziger91} and it has further been argued that it is likely to vanish at zero 
momentum in the thermodynamic limit \cite{zwanziger94}. Numerical evidence for 
this has recently been found in lattice simulations of the SU(N) gauge theory 
in $3$  \cite{cucchieri99} and 
in $4$ dimensions \cite{leinweber98,alkofer}. 
As it is well established 
that the $4d$, $SU(N)$ gauge theory at finite temperature is well described by an 
effective, $3d$ gauge theory with an adjoint Higgs field
\cite{karsch98,lacock92,kajantie97,laine99} and that, moreover, the magnetic sector 
is little influenced by the scalar Higgs fields \cite{karsch98,kajantie97,laine99}, 
the findings for the gluon propagator in $3$ dimensions are of direct relevance 
for the behavior of the magnetic gluon propagator in the $4d$ gauge theory at finite 
temperature. In fact, it recently has been suggested that also at finite temperature
the gluon propagator might be infrared suppressed \cite{zahed99}. 

It is the purpose of this letter to clarify the long-distance, low-momentum
structure of the magnetic gluon propagator of the finite temperature $SU(2)$
gauge theory in $4d$, establish the connection to the $3d$ gauge theory and
analyze to what extent these findings are gauge dependent. To this end we
will analyze the magnetic gluon propagator in Landau as well as maximally Abelian
gauges (MAG). 
Both gauges are complementary in so far as the bounds on the zero
momentum limit of the gluon propagator derived in Landau gauge
\cite{zwanziger91} are not valid in MAG because in this case the Fadeev-Popov 
matrix turns out to be quadratic in the gauge fields. 
Although details of the infrared behavior will turn out to be different 
in different gauges we show that in all cases considered
magnetic fields are exponentially screened at large distances.
In particular, we will make contact to earlier numerical calculations, which gave 
evidence for exponential damping of the magnetic propagator at large distances 
and led to the determination of a magnetic screening mass \cite{heller95}.

We have performed numerical calculations for the $4d$, $SU(2)$ gauge theory
at finite temperature as well as for the effective, $3d$ gauge theory with an  
adjoint Higgs field and the pure $SU(2)$ gauge theory in $3d$. The magnetic gluon
propagator, $D_M (z)$, and its Fourier transform, $\tilde{D}_M (p)$, have been 
calculated in Landau gauge and MAG \cite{mag} on lattices of
size $\Omega =N_\sigma^2N_z$, in $3d$ and 
$\Omega =N_\sigma^2 N_z N_\tau$ in $4d$, respectively. In our
$3d$ simulations we also have considered an anisotropic generalization of the 
Landau gauge defined by the gauge condition,
\be 
\d_1  A_1 + \d_2  A_2 +\lambda_3  \d_3  A_3=0 \quad ,
\label{l3}
\ee
which for $\lambda_3=1$ reduces to the ordinary Landau gauge condition.
For the gauge fields we use the straightforward lattice definition,
$A_{\mu}(x)  = \left[\,U_{\mu}(x)\,-\,U_{\mu}^{\dagger}(x)\,\right]/2i$,
with $U_{\mu}(x)\in SU(2)$,
which reproduces the continuum gauge fields up to ${\cal O}(a^2)$
discretization errors. We have checked previously that, up to an overall
normalization, other lattice definitions
with formally smaller discretization errors lead to identical results 
\cite{cucchieri99a}. 

The correlation function for the $\mu$-component of the gauge fields $A_{\mu}$
is defined in terms of the sum over gauge fields in a 
hyperplane orthogonal to $x_3$,
\ba
D_{\mu \mu} (z) & = &
\frac{1}{\Omega}
\biggl\langle \sum_{b=1}^3\; \sum_{x_3}\;
Q^b_{\mu}(x_3 + z)\, Q^b_{\mu}(x_3)\,\biggr\rangle\;\mbox{,}
\ea  
with 
$Q^b_{\mu}(x_3)  = 
\sum_{x_{\perp}} \, \tr [A_{\mu}(x_{\perp}\mbox{,}\,x_3)\sigma^b]/2$.
Here $\sigma^b$ are the usual Pauli matrices; 
$x_{\perp}= (x_1,x_2)$ in $3d$ and
$x_{\perp}= (x_0,x_1,x_2)$ in $4d$, respectively.
In terms of this we define the magnetic gluon propagator in
coordinate space,
\ba
D_M (z) &=& {1\over 2} (D_{11} (z) + D_{22} (z)) \quad ,
\ea
and the momentum space propagator
\be
\tilde{D}_M(p) = {\beta a^2 \over 12}
\sum_{\mu=1}^3 \; \sum_{z=0}^{N_z-1} {\rm e}^{2\pi i\;k\;z/N_z} 
D_{\mu \mu}(z)\quad ,
\ee
with $p=2\;|\sin(\pi\;k/N_z)|$ and $k=0,~1,...,~N_z-1$. Here $\beta$ is the
coupling appearing in the Euclidean lattice action and $a$ is the lattice
spacing.

The $4d$ calculations at finite temperature have been performed in the high 
temperature phase at twice the critical temperature for the deconfinement
transition. 
All calculations have been performed with the standard Wilson gauge action with  
gauge coupling $\beta=2.512$ for $N_\tau=4$ and  $\beta=2.740$ for $N_\tau=8$. These
values correspond to $T=2\;T_c$ \cite{Fingberg}.
Corresponding calculations in $3d$ have been performed at $\beta=8$ on
lattices of size  
$16^2 \times 32$, $24^2 \times 48$, $28^2 \times 56$, $32^2 \times 64$, 
$48^2 \times 64$ and $64^3$.
Moreover, we have performed $3d$ calculations at a smaller coupling,
$\beta=5$, on lattices of size $32^3$ up to $96^3$. This choice
of parameters for calculations within the reduced theory allows to
analyze the behavior of correlation functions in large 
physical volumes which are 40 times larger than what is feasible within our
$4d$ simulations.
The gauge coupling of the pure gauge theory was chosen according
to $g_3^2(T)=g^2(\mu) T$, where $g^2(\mu)$ is 1-loop running coupling constant of
the $4d$~$SU(2)$ gauge theory in the $\overline{MS}$ scheme,  $\mu=18.86 T$ and 
for $\Lambda_{\overline{MS}}$ the numerically determined relation to the deconfinement
temperature has been used, {\it i.e.} $T_c=1.06 \Lambda_{\overline{MS}}$ \cite{heller95}.
This choice of $g_3^2$ ensures a perfect description of the temperature 
dependence of the spatial string tension of the $4d$ finite temperature theory
in terms of the effective $3d$ theory \cite{bali93} down to temperatures
$T\simeq 2T_c$. Furthermore, we
have checked on our smaller lattices that the magnetic propagators calculated in
the pure $3d$~ $SU(2)$ gauge theory and in the complete $3d$ effective theory, {\it i.e.}
in the presence of an adjoint Higgs field, agree 
within statistical errors. We therefore will restrict our 
discussion here to results coming from the pure gauge sector alone. 
Our $3d$ calculations have been performed at $\beta=5$ and 8
which correspond to lattice spacings $a=0.2752/T$ and $0.172/T$, respectively.
Varying $\beta$ in our $3d$ calculations and converting to physical units with 
the above relation for $g_3^2$ we have checked that cut-off effects are small.  
To verify the cut-off independence of our results in $4d$ the calculations
have been performed at two different values of the
lattice cut-off, {\it i.e.} we used lattices of size $N_\sigma^2  N_z  N_\tau$ 
with $N_\tau=4$ and 8, $N_z = 2 N_\sigma$ and $8 \le N_\sigma \le 32$. 
Keeping the physical volume $VT^3=2(N_\sigma/N_\tau)^3$ constant we have verified 
that $D_M(z)$ is cut-off independent within the 
statistical errors of our calculation.   

\begin{figure}[t]
  \begin{center}
    \leavevmode 
\epsfig{file=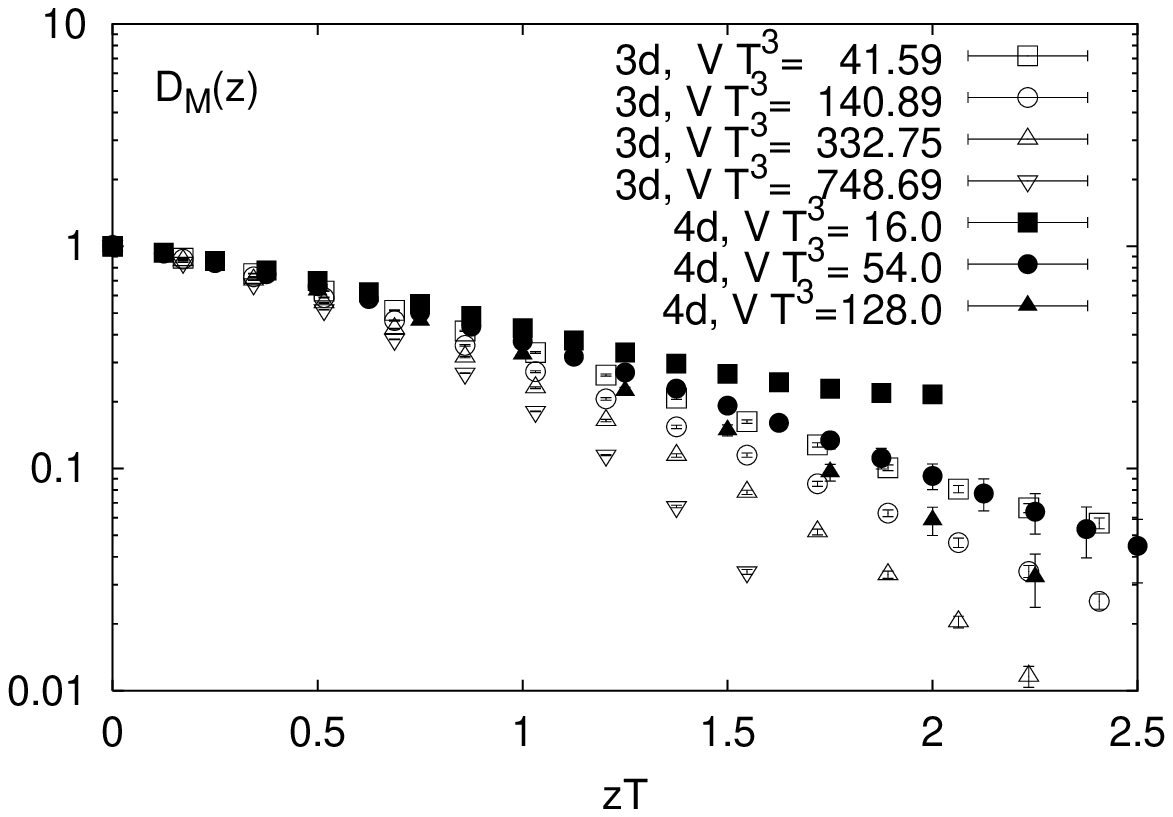,width=80mm}
  
\epsfig{file=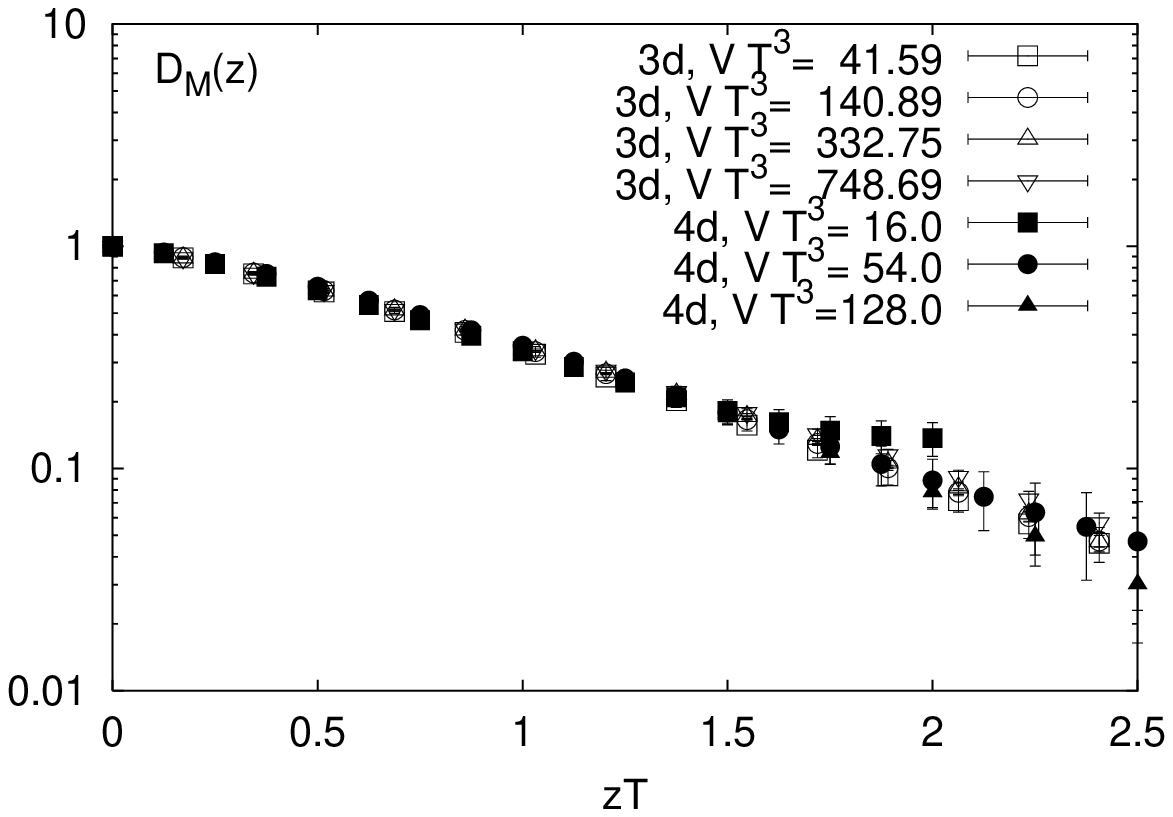,width=80mm}
\end{center}
\caption{Volume dependence of the magnetic propagators in coordinate 
space calculated in Landau (top) and maximally Abelian (bottom) gauges.
The propagators were normalized to $1$ at $z=0$.  Shown are
results from $3d$ and $4d$ simulations, which are compared at similar 
values of the physical volume in units of $T^3$. These volumes correspond to
lattices of size $16^2\times 32 \times 8$, $24^2 \times 48 \times 8$ and 
$16^2 \times 32 \times 4$ in $4d$ and to $16^2 \times 32$, $24^2 \times 48$, 
$32^2 \times 64$ and $48^2 \times 64$ in $3d$. 
} 
\label{magn_n_dep}
\end{figure}

The gluon propagator has been calculated on 1000 to 8000 gauge fixed gauge field
configurations in the case of $3d$ and about 400 configurations in $4d$ calculations.   
In Fig.~\ref{magn_n_dep} we show the magnetic propagator, $D_M(z)$, 
calculated in Landau and maximally Abelian gauges on various size lattices. 
We note that the $4d$ and $3d$ calculations performed on lattices of similar physical
size are in good agreement. 
While the magnetic gluon propagator calculated in MAG shows no significant 
volume dependence the Landau gauge propagators are strongly volume dependent 
for $zT\;\gsim\; 1$. 
A similarly strong volume dependence we also observe in
other generalized Landau gauges, {\it i.e.} gauges based on the gauge 
condition given in Eq.~\ref{l3} with $\lambda_3\ne 1$. In fact, at large distances 
the propagators in generalized Landau gauges, including the Coulomb gauge limit, become 
negative. 

\begin{figure}
  \begin{center}
    \leavevmode
\epsfig{file=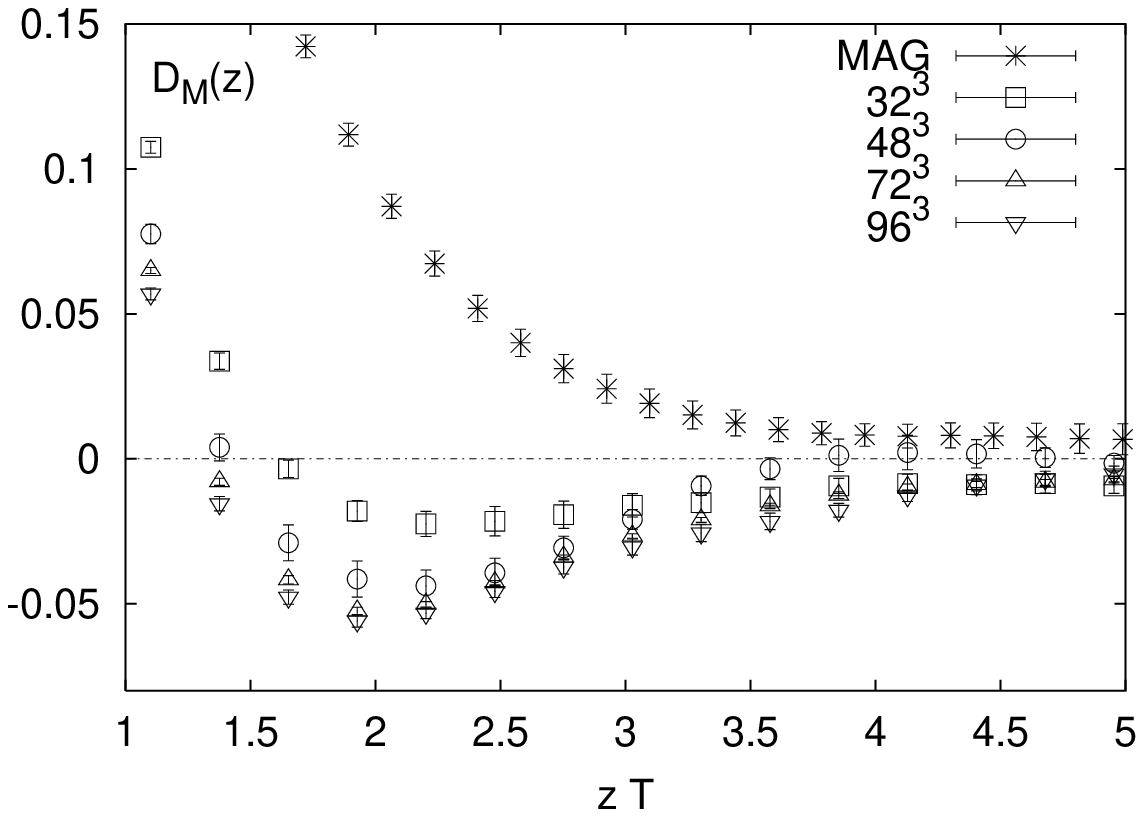,width=80mm}
\end{center}    
\caption{The magnetic gluon propagator calculated in generalized Landau gauges
with $\lambda_3= 1$ at $\beta=5$.
Also shown are data from a calculation in MAG on a 
$48^2\times 64$ lattice.
We only show the propagators at large distances. They were normalized 
to $1$ at $z=0$. 
} 
\label{demo}
\end{figure}

The virtue of the effective $3d$ theory is that we can reach here
a much larger physical volume and thus can control the volume dependence
still apparent in the comparison of $3d$ and $4d$ data shown in 
Fig.~\ref{magn_n_dep}.
We have analyzed the large distance behavior of $D_M(z)$ in these 
gauges in our $3d$ calculations on large lattices 
of size $L^3$, $L=32$, 40, 48, 56, 64, 72 and 96 at $\beta=5$. This 
corresponds to physical volumes of size $VT^3=683$ up to $VT^3=5464$.
In all cases we clearly observe negative correlation functions 
at large distances. For some of these lattices we show the long-distance
part of the $D_M(z)$ in Fig.~\ref{demo} for $\lambda_3 = 1$.
Similar results we find for $\lambda_3 \ne 1$.
We note that the volume dependence is small on these large lattices.
Moreover, the correlation functions become smaller with increasing
lattice size. This is in accordance with our expectation that finite 
size effects are due to zero mode contributions which disappear
in the infinite volume limit. On finite lattices, however, they add a 
volume-dependent, positive constant to the correlation function in 
coordinate space.

Our analysis thus gives clear evidence that the magnetic gluon
correlation function in coordinate space calculated in generalized 
Landau gauges becomes negative at large distances, {\it i.e.} for 
$zT\gsim 1.5$.
At $T=0$ such a behavior is, in fact, expected to occur for the gluon 
propagator as a consequence of the conjecture made by
Zwanziger that the Landau gauge gluon propagator vanishes in the
infrared limit. In coordinate space it has been suggested that
the propagator takes on the form
\cite{zwanziger91}
\be
D_M(z) = A {\rm e}^{-m_M z}\; \cos(b m_M z + c) \quad ,
\label{ansatz}
\ee
Although we can at present not quantify the functional form
of the correlation function in generalized Landau gauges, 
our results are consistent with this ansatz up to distances
$zT \simeq 3$.
To judge whether the correlation function will indeed
continue to oscillate as suggested by Eq.~\ref{ansatz} will, 
however, require a more detailed analysis of the 
large distance behavior of $D_M (z)$.
In general, for $\tilde{D}_M(p)$ to vanish
at $p=0$ in momentum space one should find $\tan (c) = 1/b$ with the 
above ansatz. This conjecture, $\lim_{p\rightarrow 0} \tilde{D}_M(p) =0$,
may also be valid at non-zero temperature \cite{zahed99}. 

In contrast to the complicated structure of the Landau gauge propagator
the propagator calculated in MAG does show a simple exponential decay at
large distances and
does not show any significant volume dependence. We could 
verify that it stays positive at least up to distance $zT=3.5$.
For $zT>1.5$ simple exponential fits gave $\chi^2/(d.o.f) = 1.7$ and we 
also found that local masses became independent of $z$ within statistical
errors for $zT>1$. Best fits with the ansatz given in Eq.~\ref{ansatz}
were thus obtained for $b=0$. From this we find for the magnetic screening
mass of the magnetic propagator calculated in MAG,
\be
{m_M\over T}  = 1.48\pm 0.17 ~~{\rm at}~~T=2 T_c\quad.
\label{mfit}
\ee
When taking the absolute value of the Landau gauge propagator this is
found to be bounded by the exponentially decaying MAG-propagator. This
confirms that also in Landau gauge correlations of magnetic fields
are exponentially screened.

\begin{figure}
  \begin{center}
    \leavevmode 
\epsfig{file=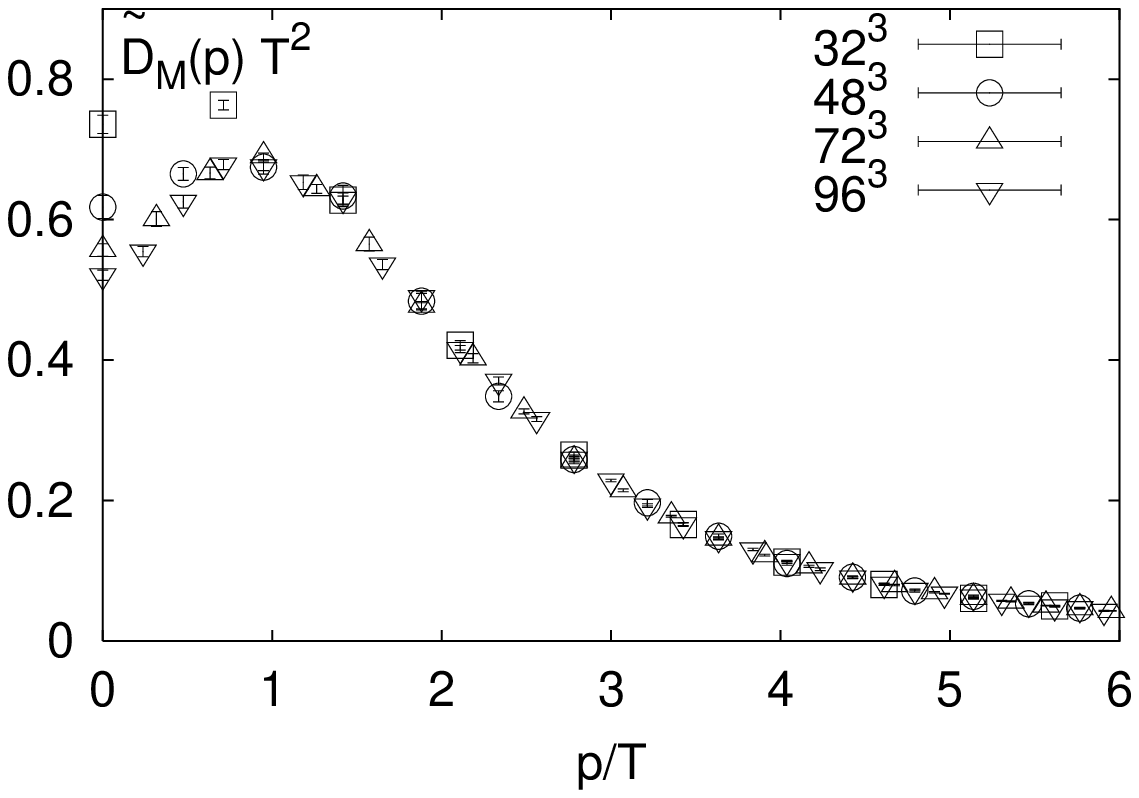,width=80mm}
\end{center}
\caption{The magnetic gluon propagator in momentum space calculated in 
Landau gauge on $3d$ lattices of various sizes at $\beta=5$. 
}
\label{magn_p}
\end{figure}

The fact that the magnetic gluon propagator calculated in Landau gauge
becomes negative at large distances translates into a suppression  
of the momentum space propagator at small momentum. This is shown in  
Fig.~\ref{magn_p} for the Landau gauge propagators, similar results hold
for $\lambda_3\ne 1$. 
For $p<T$ the  momentum space propagator is  
sensitive to the volume, while for large momenta ($p > T$) it is essentially 
independent of the lattice size.   
Moreover, we note that for small momenta the finite volume effects lead 
to a decrease of $\tilde{D}(p)$ with increasing volume while the volume
dependence is already negligible for $p/T \simeq 1$. The occurrence of
a maximum in $\tilde{D}(p)$ for $p/T \simeq 1$ and the infrared
suppression of the momentum space propagator at $p=0$ thus is firmly
established by our data.

As expected the volume dependence of the magnetic propagator in
momentum space 
is strongest at $p=0$. On large lattices  $\tilde{D}_M(p)$ reaches a  
maximum at a non-zero but small value of $p/T$ and keeps decreasing 
at $p=0$ with increasing lattice size. This 
might indicate that $\tilde{D}_M(0)$ could actually vanish in the infinite 
volume limit. 
The propagator at vanishing 
momentum reflects the contribution of zero momentum fluctuations of the 
gauge fields,
\be
\tilde{D}_{M}(0) = {\beta a^2\over 12} \Omega 
\sum_{\mu, b} \langle {(\phi_{\mu}^b)}^2 \rangle {\rm ~with~} 
\phi_{\mu}^b  = {1\over \Omega} \sum_x A_{\mu}^b(x)~.
\ee
In order for it to vanish in the infinite volume limit the zero 
mode fluctuations $\langle {(\phi_{\mu}^b)}^2 \rangle$ thus have to 
drop faster than $\Omega^{-1}$. 
To quantify this behavior 
we have fitted  $\tilde{D}_M (0) T^2$ to a simple ansatz that assumes
a power-like dependence on the volume,
$\tilde{D}_M (0)T^2 =  b (VT^3)^{-z} + c $.
At present such fits to our $4d$ and $3d$ data can, however,
not rule out a
non-vanishing value for the constant $c$. We stress, however,
that the value of $\tilde{D}_M (0)$ itself is of little importance
for the observed complex structure of the gluon propagator in Landau 
gauge in coordinate as well as in momentum space and in any case is
gauge dependent. 
Nonetheless,
the behavior found here for the finite-T magnetic gluon propagator is consistent
with findings of Ref. \cite{cucchieri99} for the $T=0$ gluon propagator of the
$3d$, $SU(2)$ gauge theory. There it was shown that the magnetic propagator 
calculated in Landau gauge is less singular then $p^{-1}$ in the infrared and is
likely to vanish in $p \rightarrow 0$, $V\rightarrow \infty$ limit. 

The momentum dependence of $\tilde{D}_M(p)$ shown in Fig.~\ref{magn_p} also 
makes clear why earlier calculations on medium size lattices let to the 
determination of a non-zero magnetic mass. The infrared suppression becomes
significant only for momenta $p/T \lsim 0.5$.  
The existence of a simple {\it pole mass} becomes unlikely on the basis of 
our analysis. However, we stress that magnetic fields in coordinate space 
are, of course, strongly screened. The magnetic gluon propagator 
calculated in Landau gauge as well as MAG is exponentially
damped.  
The value for the screening mass extracted from the propagator in MAG gauge
is in good agreement with earlier in $4d$ \cite{heller95} and $3d$ \cite{karsch98}
lattice calculations in Landau gauge, although the data analysis in these cases 
was based on a simple pole ansatz for the magnetic mass.

In the course of the calculations reported here we also have reanalyzed the
behavior of the electric gluon propagator in different gauges. In that case
no suppression in the infrared has been observed and the electric screening mass has 
been found to be gauge independent within statistical errors. Details of this
calculation as well as a more detailed discussion of our results for the 
magnetic gluon propagator will be presented elsewhere.

\vspace{0.5cm}
\noindent
{\bf Acknowledgements:}

\medskip
\noindent
The work has been supported by the TMR network
ERBFMRX-CT-970122 and by the DFG under grant Ka 1198/4-1.
Our calculations have partly been performed at the HLRS in 
Stuttgart and the $(PC)^2$ in Paderborn.

\end{document}